\documentclass[twocolumn,aps,pra]{revtex4-1}
\usepackage{epsfig}
\usepackage{amsmath}
\usepackage{amsfonts}
\usepackage{color}
\usepackage{soul}
\usepackage{multirow}
\usepackage{amssymb}

\def\a{\hat{c}}
\def\aav{\left \langle \a \right\rangle}
\def\apv{\left \langle \a^+ \right\rangle}

\def\B{\Bar{B}}

\def\dnuD{\Delta \nu_{\rm D}}
\def\dnuS{\Delta \nu_{\rm S}}
\def\dnuSs{\Delta \nu_{\rm S,0}}
\def\dnuSt{\Delta \nu_{\rm S,2}}

\def\g{g_j}

\def\E{\Bar{E}}
\def\EE{\mathcal{E}}
\def\EEE{\Hat{\Bar{E}}}

\def\F{\mathcal{F}}

\def\H{\hat{H}}
\def\HH{\hat{\mathcal{H}}}

\def\I{\Bar{\Bar{I}}}
\def\II{\Hat{\Bar{I}}}

\def\JJ{\Hat{\Bar{J}}}

\def\k{\Bar{k}}

\def\LL{\hat{\hat{\mathcal{L}}}}

\def\Lamrf{\Bar{\Bar{\Lambda}}_{\rm rf}}
\def\Lams{\Bar{\Bar{\Lambda}}_{\rm s}}

\def\rperp{r_\perp}

\def\Q{\Bar{\Bar{Q}}}
\def\QQ{\Hat{\mathcal{Q}}}

\def\r{\Bar{r}}
\def\R{\Bar{R}}

\def\Veff{V_{\rm eff}}
\def\w{w_0}

\def\hro{\hat{\varrho}}

\def\sig{\hat{\sigma}}

\def\sigll{\langle \sig^j_{ll}\rangle}
\def\siguu{\langle \sig^j_{uu}\rangle}
\def\siglu{\langle \sig^j_{lu}\rangle}
\def\sigul{\langle \sig^j_{ul}\rangle}

\def\sigii{\langle \sig^j_{ii}\rangle}
\def\sigil{\langle \sig^j_{il}\rangle}
\def\sigli{\langle \sig^j_{li}\rangle}
\def\sigui{\langle \sig^j_{ui}\rangle}
\def\sigiu{\langle \sig^j_{iu}\rangle}
\def\sigz{\langle \sig^j_{z}\rangle}

\def\e{\Bar{\mathrm{e}}}

\begin{document}

\title{Prospects for a bad cavity laser using a large ion crystal}
\author{Georgy~A.~Kazakov$^1$\thanks{E--mail:
kazakov.george@gmail.com}, Justin~Bohnet$^2$, Thorsten~Schumm$^1$}
\affiliation{{\setlength{\baselineskip}{18pt}
{$^1$Vienna University of Technology, Atominstitut, Stadionallee 2, 1020 Vienna, Austria}\\
{$^2$National Institute of Standards and Technology, Time and Frequency Division, Boulder, Colorado 80305, USA}\\
}}


\begin{abstract}
We propose to build a bad cavity laser using forbidden transitions in large ensembles of cold ions that form a Coulomb crystal in a linear Paul trap. This laser might realize an active optical frequency standard able to serve as a local oscillator in next-generation optical clock schemes. In passive optical clocks, large ensembles of ions appear less promising, as they suffer from inhomogeneous broadening due to quadrupole interactions and micromotion-relates shifts. In bad cavity lasers however, the radiating dipoles can synchronize and generate stable and narrow-linewidth radiation. Furthermore, for specific ions, micromotion-induced shifts can be largely suppressed by operating the ion trap at a magic frequency. We discuss the output radiation properties and perform quantitative estimations for lasing on the ${^3D_2} \rightarrow {^1S_0}$ transition in ${\rm ^{176}Lu^+}$ ions in a spherically-symmetric trap.
\end{abstract}


\maketitle                

\section{Introduction}
\label{sec:intro}
Optical frequency standards are the most stable clocks to date. The most advanced implementations reach a short-term stability at the $3.4\times 10^{-16}/\sqrt{\tau}$ level \cite{Bloom14}, and systematic uncertainty of $3.2\times 10^{-18}$ \cite{Huntemann16}. Further improvement of optical frequency standards would allow a multitude of new applications in fundamental and applied science, such as study of fundamental constant variations \cite{Godun14} and relativistic geodesy \cite{Chou10}. Modern optical clocks are {\em passive clocks}, where the frequency of a {\em local oscillator}, i.e., some stable narrow-band laser, is feedback-stabilized to a narrow and robust {\em etalon} transition in trapped atoms or ions. This etalon transition may be extremely narrow, down to a nHz level in some species \cite{Huntemann16}, but the real spectroscopic linewidth is limited by the short-term stability of the local oscillator and usually does not surpass the sub-Hz level. Also, on a timescale shorter than the interrogation time of the etalon transition, the stability of the local oscillator entirely determines the stability of the whole frequency standard. Fluctuations of the local oscillator frequency may also contribute to the instability of the frequency standard on longer timescales via the Dick effect \cite{Audoin98}. Therefore, improving the local oscillators is one of the key tasks for the development of more precise optical clocks. The best modern local oscillators are lasers that are prestabilized to an ultrastable macroscopic cavity. Their stability is usually limited by mechanical and thermal noise \cite{Numata04}, and may attain a level of $8\times 10^{-17}$ on the timescale up to $10^3$~s at room temperature \cite{Haefner15}, and even $4\times 10^{-17}$ on the timescale up to $10^2$~s in cryogenic environments \cite{Matei17}, but the progress in this direction is slow.

One possible alternative approach is to create an {\em active optical frequency standard}, i.e., a laser where atoms with a narrow and robust lasing transition play the role of the gain medium. Such a laser would operate in the so-called {\em bad cavity} regime, where the linewidth of the cavity mode is much broader than the gain profile. The output frequency of such a laser is determined primarily by the gain medium, which makes it robust to fluctuations of the cavity length. Such standards have been proposed by several authors recently \cite{Chen05, Meiser09, Yu08}, and a series of proof-of-principle experiments have been performed \cite{Bohnet12, Cox14, Weiner15, Norcia15, Norcia16}. 

Active atoms that constitute the gain for an active optical frequency standard must be confined to the Lamb Dicke regime to avoid Doppler and recoil shifts. Such a confinement may be realized with an optical lattice potential at a so-called {\em magic} frequency, where the upper and the lower lasing states experience the same ac Stark shift \cite{Derevianko11}. These shifts depend on the polarization of the trapping fields and can be controlled to the necessary level of precision only for ${^3P_0} \rightarrow {^1S_0}$ transitions in Sr and other alkali-earth atoms, Zn, Cd, Hg and Yb. A first proof-of-principle experiment with such a transition in trapped Sr atoms has been recently performed in a pulsed regime \cite{Norcia16}.

The optical lattice potential trapping neutral atoms is relatively shallow, of order of a few tens of $\mu$K \cite{Derevianko11}. This leads to a short trap lifetime, therefore some method of compensating for atom losses must be implemented to practically realize an active optical frequency standard \cite{Kazakov13, Kazakov14}. The implementation of such methods is rather complicated, although certain efforts in this direction are being made \cite{Schreck}.

In contrast to neutral atoms, charged ions may be trapped in much deeper Paul or Penning traps, which leads to much longer trap lifetimes. Trapped ions may also be cooled via co-trapped ions of another species (sympathetic cooling) \cite{Bowe99}. A bad cavity laser utilizing trapped ions may operate continuously over hours, even days, without the need to compensate for ion losses. On the other hand, micromotion of the ions and their interactions with trapping fields and with each other causes shifts and inhomogeneous broadening of the etalon transition; these effects are especially pronounced in large ion ensembles. Thus, ion optical clocks have been built primarily with single ions \cite{Huntemann16, Godun14, Chou10} or with few-ion ensembles \cite{Keller16}. 

Inhomogeneous broadening may be considerably reduced for ions with negative differential polarizability of the clock states in RF Paul traps at a specially chosen {\em magic} frequency of the trapping field \cite{Arnold15}. Also, bad cavity lasers with inhomogeneously broadened gain may produce synchronous and stable output radiation, if the total {\em homogeneous} broadening of the lasing transition exceeds the inhomogeneous broadening by at least a few times \cite{Kazakov17}. In lasers based on a 3-level scheme, such homogeneous broadening will be dominated by repumping, which opens the possibility to build a bad cavity laser with ions \cite{Kazakov17}.


In this paper, we present a detailed discussion of the bad cavity laser based on Coulomb crystals in Paul traps. In Section~\ref{sec:micro} we consider a generic model of a harmonic co-axial Paul trap formed by static and radio-frequency (rf) harmonic potentials, obtain general expressions for micromotion-induced Doppler and Stark shifts in a cold Coulomb crystal, and introduce the ``magic'' frequency, which allows one to compensate these shifts in leading order. In Section~\ref{sec:res} we consider residual terms of the shifts, and specify the trap geometry. In Section~\ref{Sec:Calc} we derive the equation for the intracavity field, taking into account standing-wave periodicity and Gaussian shape of the cavity mode. In Section~\ref{sec:Lu} we present some quantitative estimations for a bad cavity laser with trapped ${\rm ^{176}Lu^+}$ ions. In Section~\ref{sec:con} we discuss the results, envisaged difficulties, and possible ways to overcome them.

\section{Magic frequency}
\label{sec:micro}
Here we consider micromotion-induced second-order Doppler and dc Stark shifts for ions forming a cold Coulomb crystal in a harmonic rf Paul trap. Such a many-ion crystal has been considered in \cite{Arnold15}, although some higher-order terms have been omitted there. These terms, however, can be easily calculated, if we note that the only macroscopic force acting on the ion in the Paul trap is proportional to the same local electric field that causes the Stark shift.

We consider a Paul trap formed by the potential
\begin{equation}
\begin{split}
\phi (\r, t)&=\frac{m \Omega \omega_z}{2 q} \, \r^T \cdot \left[ \Lamrf \cos (\Omega t)+\frac{\epsilon}{2} \Lams \right] \cdot \r,
\label{equation:1}
\end{split}
\end{equation}
where $\r=x \e_x+y \e_y+ z\e_z\equiv x_1 \e_x+x_2 \e_y+ x_3 \e_z$ is the position vector, $m$ and $q$ are the mass and the charge of the ion, $\Omega$ is the rf drive frequency, $\omega_z$ the frequency characterizing the trap confinement, $\epsilon=2 \omega_z/\Omega$, and $\Lamrf$ and $\Lams$ are traceless dimensionless symmetric matrices determining curvatures of the potentials. We use a single bar to denote column vectors with 3 spatial components, a double bar to denote $3\times 3$ matrices, a dot ($\cdot$) for the inner product, and the superscript $T$ for the transposition (we will often omit this superscript for vectors, for the sake of brevity). Also we suppose that $\omega_z \ll \Omega$, i.e., such that $\epsilon$ may be considered as a small parameter.

Scaling time and length (in Gaussian units) by
\begin{equation}
\frac{t\Omega}{2} \rightarrow t, \quad \frac{\r}{\ell} \rightarrow \r, \quad {\rm where} \quad 
\ell=\left(\frac{q^2}{m \omega_z^2} \right)^{1/3}, 
\label{equation:2}
\end{equation}
we write the equation of motion (e.o.m.) of $i$-th ion as
\begin{equation}
\ddot{\r}_i+\epsilon^2(\Lams \cdot \r_i)+2 \epsilon (\Lamrf \cdot \r_i) \cos(2 t)=\epsilon^2 \sum_{j\neq i} \frac{\r_{ij}}{r^3_{ij}}\, ,
\label{equation:3}
\end{equation}
where $\r_{ij}=\r_i-\r_j$, $r=|\r|$.
Following \cite{Arnold15, Landa12} we assume the existence of a stable $\pi$-periodic solution of the e.o.m. (\ref{equation:3}), which may be expressed as
\begin{equation}
\r_i(t) = \R_{0,i}+2 \sum_{n=1}^{\infty} \R_{2n,i} \cos(2 n t).
\label{equation:4}
\end{equation}
We suppose that all the motions of the ions except the micromotion (\ref{equation:4}) are frozen out (cold Coulomb crystal).

The main trap-induced corrections to the frequency $\nu=2 \pi \omega$ of the clock transition of the $i$th ion are the micromotion-induced second-order Doppler shift $\dnuD^i$, and the Stark shift $\dnuS^i$ caused by the time-dependent local electric field of the trap and nearby ions acting on the $i$th ion at its instantaneous position. The second-order fractional Doppler shift averaged over the period of the micromotion is
\begin{equation}
\frac{\dnuD^i}{\nu}=
-\frac{\Omega^2 \ell^2}{4}\frac{\langle \dot{\r}^2 \rangle}{2 c^2}=-\frac{\Omega^2 \ell^2}{c^2} \sum_{n=1}^{\infty}n^2 \R_{2n,i}^2,
\label{equation:5}
\end{equation}
where $c$ is the speed of light, and the prefactor comes from the scaling (\ref{equation:2}).

We first consider the Stark shift of the clock transition caused by the local electric field acting on the ion. Following \cite{Itano00}, we suppose that the ion trap is placed into a homogeneous external magnetic field $\B$ causing the Zeeman splitting to be much larger than the tensor component of the Stark shift (see estimations in the end of Sec.~\ref{sec:Lu}). Then the Stark shift of some Zeeman sublevel can be written as 
\begin{equation}
\Delta \EE = -\frac{\alpha_0}{2} \E^2 - \frac{\alpha_2}{4} (3 E_z^2-\E^2),
\label{equation:6}
\end{equation}
where the axis $z$ is oriented along $\B$, $\alpha_{0}$ is the scalar polarizability, and $\alpha_{2}$ is
\begin{widetext}
\begin{eqnarray}
\alpha_2&=&\alpha_2(\eta,J,F,m_F)= \alpha_{tens} (\eta,J) \frac{3m_F^2-F (F+1)}{3F^2-F (F+1)} (-1)^{I+J+F}
 \left\{
\begin{array}{ccc}
F & J & I \\
J & F & 2
\end{array}
 \right\} \nonumber \\
 & & \times \left( 
\frac{F (2F-1) (2F+1)(2J+3)(2J+1)(J+1)}{(2F+3)(F+1)J(2J-1)}
\right)^{1/2}. \label{equation:7}
\end{eqnarray}
\end{widetext}
Here $\eta, J, F, m_F$ are the principal quantum number, the angular momentum of the electronic shell, the total angular momentum and its projection onto the direction of the magnetic field respectively, and $\alpha_{tens} (\eta,J)$ is the tensor polarizability characterizing the state $\eta, J$ of the electronic shell of the atom \cite{Itano00}. 

Denoting the differential polarizabilities $\Delta \alpha_{k} = \alpha_{k}^u - \alpha_{k}^l$ ($k=0$ or 2) of the upper ($u$) and lower ($l$) clock states, and taking into account the relation between the local electric field and the instantaneous acceleration of the ion, we can write the time-averaged scalar and tensor Stark shifts $\dnuSs^i$ and $\dnuSt^i$ of the clock transition of the $i$th ion as

\begin{align}
\dnuSs^i &= -\frac{\Delta \alpha_0}{4 \pi \hbar} \langle \E^2 \rangle= -\frac{\Delta \alpha_0}{4 \pi \hbar} \frac{\Omega^4 \ell^2 m^2}{16 q^2} \langle \ddot{\r}^2 \rangle \nonumber \\
        &= - \frac{\Delta \alpha_0}{2 \pi \hbar} \frac{\Omega^4 \ell^2 m^2}{q^2} \sum_{n=1}^{\infty} n^4 \R^2_{2n,i}, \label{equation:8}
\\
\dnuSt^i &=\frac{\Delta \alpha_2}{8 \pi \hbar} (\langle \E^2 \rangle - 3 \langle \E_z^2 \rangle)  
\nonumber \\
        & = \frac{\Delta \alpha_2}{4 \pi \hbar} \frac{\Omega^4 \ell^2 m^2}{q^2}    
\sum_{n=1}^{\infty} n^4 (\R^2_{2n,i}-3 Z^2_{2n,i}). \label{equation:9}  
\end{align}
where $Z_{2n,i}$ is the $z$-projection of $R_{2n,i}$.

Combining (\ref{equation:5}), (\ref{equation:8}) and (\ref{equation:9}), we can express the sum of the Doppler and Stark shifts in the form
\begin{multline}
\Delta \nu^i =\dnuS^i+\dnuD^i =\\
 -\frac{\Omega^2 \ell^2 \nu}{c^2} 
	\left[
		 1 + \left(
		 	       \Delta \alpha_0-\frac{\Delta \alpha_2}{2}
		 	 \right)
		 	 \frac{(m \Omega c)^2}{2 \pi \hbar \nu q^2}
	\right]
    \sum_{n=1}^{\infty} n^4 \R^2_{2n,i} \\
+
\frac{\Omega^2 \ell^2 \nu}{c^2} \sum_{n=1}^{\infty}(n^4-n^2) \R^2_{2n,i} \\
-
3 \Delta \alpha_2 \nu \frac{\Omega^4 m^2 \ell^2}{4 \pi \hbar \nu q^2} \sum_{n=1}^{\infty} Z^2_{2n,i} n^4. \label{equation:10}
\end{multline}
It is easy to see that if $\Delta \alpha_2 > 2 \Delta \alpha_0$, the first term in (\ref{equation:10}) will be zero at so-called {\em magic} value $\Omega_0$ of the radio frequency $\Omega$:
\begin{equation}
\Omega_0=\frac{q}{mc} \sqrt{\frac{4 \pi \hbar \nu}{\Delta \alpha_2-2 \Delta \alpha_0}}.
\label{equation:11}
\end{equation}
In the next section we consider remaining terms of (\ref{equation:10}).

\section{Residual micromotion-related shifts}
\label{sec:res}
\begin{figure}
\begin{center}
\resizebox{0.45\textwidth}{!}
{\includegraphics{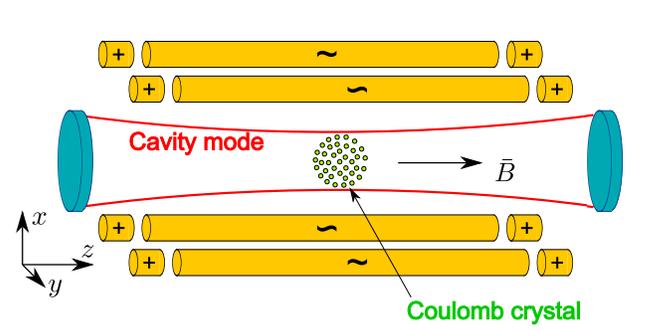}}
\end{center}
\caption{Sketch of the linear Paul trap with Coulomb crystal and external cavity (not to scale), where the $|F_u, m_u=0\rangle \rightarrow |F_l, m_l=\pm 1\rangle$ quadrupole transitions are coupled with two circularly polarized cavity modes. $\B$ is the magnetic field determining the quantization axis.
\label{fig:f1}}
\end{figure}
To estimate residual terms in (\ref{equation:10}), we expand $\R^2_{2n,i}$ and $Z^2_{2n,i}$ by the small parameter $\epsilon=2\omega_z/\Omega$. Also we suppose that the oscillating terms $\R_{2n,i}$~($n\neq 0$) are small in comparison with the time-independent components $R_{0,i}$. Then we can decompose the right part of the e.o.m. (\ref{equation:3}) as
\begin{equation}
\frac{\r_{ij}}{r_{ij}^3}=\frac{\R_{0,ij}}{R_{0,ij}^3}+\Q_{ij} \cdot \r_{ij}^{\prime}+..., \label{equation:12}
\end{equation}
where $\r_{ij}^{\prime}=\r_{ij}-\R_{0,ij}$, and 
\begin{equation}
\Q_{ij}=-\frac{3 \R_{0,ij} \otimes \R_{0,ij} - \I R_{0,ij}^2}{R_{0,ij}^5}. 
\label{equation:13}
\end{equation}
Here the symbol $\otimes$ denotes the outer product, and $\I$ is the identity matrix. 

Substituting (\ref{equation:13}) and (\ref{equation:4}) into (\ref{equation:3}), we obtain:
\begin{align}
\R_{2,i}=&\frac{\epsilon}{4} \Lamrf \cdot \R_{0,i}+\frac{\epsilon^3}{16}
          \left[
               \left(\Lams+\frac{\Lamrf^2}{16} \right) \cdot \Lamrf \cdot \R_{0,i} 
          \right. \nonumber \\
         &\left. \vphantom{\left(\Lams+\frac{\Lamrf^2}{16} \right)}     
               -\sum_{j\neq i} \Q_{ij}\cdot \Lamrf \cdot \R_{0,ij}
          \right]+O(\epsilon^5), 
\label{equation:14} 
\\
\R_{4,i}=&\frac{\epsilon^2}{64} \Lamrf^2 \cdot \R_{0,i}+O(\epsilon^4), 
\label{equation:15} 
\\
\R_{6,i}=&\frac{\epsilon^3}{2304} \Lamrf^3 \cdot \R_{0,i}+O(\epsilon^5).
\label{equation:16}
\end{align}

Consider the two residual terms in (\ref{equation:10}). The second term contains only the summands with $n\geq2$. It may be estimated as
\begin{equation}
\begin{split}
\frac{\Omega^2 \ell^2}{c^2} \sum_{n=1}^{\infty}(n^4-n^2) \R^2_{2n,i}\approx  
12 \frac{\Omega^2 \ell^2}{c^2} \R_{4,i}^2 \\
\approx \frac{3 \epsilon^4}{1024} \frac{\Omega^2 \ell^2}{c^2} \R_{0,i}\cdot \Lamrf^4 \cdot \R_{0,i}.
\label{equation:17}
\end{split} 
\end{equation}
The last term of (\ref{equation:10})
\begin{equation}
- 3 \Delta \alpha_2\frac{\Omega^4 m^2 \ell^2}{4 \pi \hbar \nu q^2} \sum_{n=1}^{\infty} Z^2_{2n,i} n^4
\label{equation:18}
\end{equation}
contains also the summand with $n=1$, which is proportional to $\epsilon^2$. However, it can be substantially reduced by a proper choice of the trap geometry. Namely, if the radio frequency component of the trap field is orthogonal to the $z$ axis (see Fig.~\ref{fig:f1}), i.e., if
\begin{equation}
\Lamrf=
\left[
\begin{array}{ccc}
a & 0 & 0 \\
0 & -a & 0 \\
0 & 0 & 0
\end{array}
\right], 
\label{equation:19}
\end{equation}
then $Z_{6,i}=O(\epsilon^4)$, $Z_{4,i}=O(\epsilon^4)$, and $Z_{2,i}=O(\epsilon^3)$. Therefore, the whole last term of (\ref{equation:10}) is of order of $\epsilon^6$, and it is dominated by the second term of (\ref{equation:10}). We can neglect it, and approximate $\Delta \nu^i$ as
\begin{equation}
\Delta \nu^i
\approx \frac{3 \nu \epsilon^4}{1024} \frac{\Omega^2 \ell^2}{c^2} a^4 (X_{0,i}^2+Y_{0,i}^2).
\label{equation:20}
\end{equation}

For our future estimations we consider the particular case of a spherical trap. Namely, we take $a=\sqrt{3}$ in (\ref{equation:19}), and 
\begin{equation}
\Lams=
\left[
\begin{array}{ccc}
-\frac{1}{2} & 0 & 0 \\
0 & -\frac{1}{2} & 0 \\
0 & 0 & 1
\end{array}
\right], 
\label{equation:21}
\end{equation}
which corresponds to the pseudopotential 
\begin{equation}
V(\r)=\frac{m \omega_z^2}{2} \, \r \cdot \left(\Lams+\frac{\Lamrf^2}{2} \right)\cdot \r
=\frac{m \omega_z^2 \r^2}{2}.
\label{equation:22}
\end{equation}
It is easy to show that a large Coulomb crystal of $N$ ions in such a trap will have an approximate spherical shape with the radius $R\approx N^{1/3}$ in the units of $\ell$ (i.e., $\ell$ is the {\em Wigner-Seitz radius}), and the density of the crystal will be homogeneous on the scales exceeding $\ell$. 

The micromotion-related shift (\ref{equation:20}) goes to zero at the center of the crystal and reaches its maximum value 
\begin{equation}
\Delta_{max} = 2 \pi \, \Delta \nu^i_{max} \approx \left(\frac{3}{4} \right)^3 
\frac{2 \pi \nu \,N^{2/3} \, \omega_z^{8/3}\, q^{4/3}}{\Omega_0^2 \, c^2\, m^{2/3}}
\label{equation:23}
\end{equation}
at $X_{0,i}^2 + Y_{0,i}^2 = N^{2/3}$. 

\begin{figure}
\begin{center}
\resizebox{0.45\textwidth}{!}
{\includegraphics{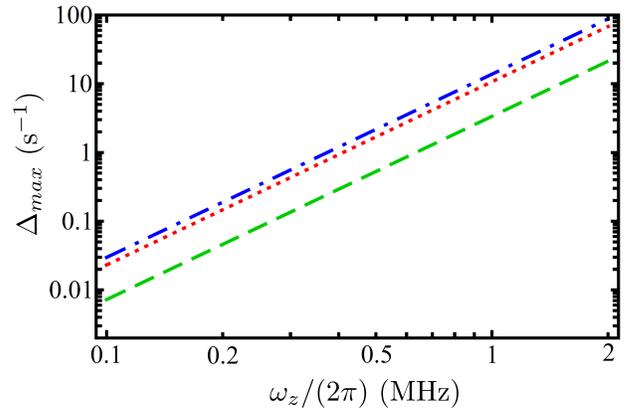}}
\end{center}
\caption{Maximum micromotion-induced shift $\Delta_{max}$ calculated according (\ref{equation:23}) at $N=10^5$ ions for different $|{^3D_2}, F_u,m_F=0\rangle \rightarrow |{^1S_0}, F_l=7,m_F=1\rangle$ transitions in ${\rm ^{176}Lu^{+}}$ ions. Red dotted line: $F_u=5$, $\Omega=2\pi \times 25.3$~MHz; green dashed: $F_u=8$, $\Omega=2\pi \times 55.3$~MHz; blue dot-dashed: $F_u=9$, $\Omega=2\pi \times 22.3$~MHz.}
\label{fig:shift}
\end{figure}

To illustrate the dependence of $\Delta_{max}$ on $\omega_z$, we present in Fig.~{\ref{fig:shift}} the maximal micromotion-related shifts $\Delta_{max}(\omega_z)$ for different $|{^3D_2},F_u,m_F=0\rangle \rightarrow |{^1S_0},F_l,m_F=1\rangle$ transitions in ${\rm ^{176}Lu^{+}}$ ions, supposing that the spherical Coulomb crystal contains $N=10^5$ ions, and that the radio-frequencies $\Omega$ are equal to the magic frequencies $\Omega_0$ for the corresponding transitions, see Section~{\ref{sec:Lu}} for details.

We should note that the expressions (\ref{equation:20}) -- (\ref{equation:23}) for the micromotion-related shift as well as for the magic frequency $\Omega_0$ (\ref{equation:11}) of the radio-frequency field have been obtained in the leading order ($\epsilon^4$) of the small parameter $\epsilon$, and in this leading order the individual shift $\Delta_i \propto X_i^2+Y_i^2$. However, it is easy to see that the first term in (\ref{equation:10}) (which turns to zero at $\Omega=\Omega_0$) is also proportional to $X^2+Y^2$ in the leading order of $\epsilon^2$. Therefore, fine tuning of $\Omega$ near $\Omega_0$ may be used for further compensation of the micromotion-related shifts down to the order of $\epsilon^6$; the respective correction of the magic frequency has been considered in \cite{Arnold15}. Also, this tuning may be used for compensation of the light shifts caused by the pumping and cooling fields, or for suppression of the sensitivity of the frequency of the lasing transition to fluctuations of non-perfectly controlled parameters of the trap, such as amplitudes of the trapping and pumping fields. Detailed investigations of these possibilities are beyond the scope of this paper. 

\section{Cavity field}
\label{Sec:Calc}
In this section we estimate the output power of the bad cavity laser based on a spherical Coulomb crystal with radius $R_c=N^{1/3} \ell$ coupled with the cavity field. We neglect here the micromotion-induced and quadrupole shifts of the lasing transitions; this assumption is acceptable, if the inhomogeneous broadening caused by these shifts is small in comparison with the homogeneous one \cite{Kazakov17}. These assumptions will be proven in the end of Section~\ref{sec:Lu}. Instead, we take into account that the cavity mode is a standing-wave Gaussian mode with waist $w_0$. 

We start from the mean-field equations (see Appendix~\ref{app:II} for details of the derivation), where we neglect detunings and suppose equivalence of the cavity eigenfrequency $\omega_c$ with the transition frequencies $\omega$ and $\omega_{ul}^j$ of the laser field and lasing transitions:
\begin{align}
&\frac{d\aav}{dt}  =- \frac{\kappa}{2} \aav 
                     -\frac{i}{2}\sum_j \g \siglu, \label{equation:24} 
\\
&\frac{d \sigz}{dt}=i\g \left[\apv \siglu - \aav \sigul \right]
                   -\gamma_\parallel \sigz + w - \gamma, \label{equation:25}
\\
&\frac{d \siglu}{dt}=\frac{i\g}{2}\aav \sigz - \gamma_\perp \siglu. \label{equation:26}
\end{align}
Here $\a$, $\a^+$ are the cavity field operators, $\sig_{\alpha \beta}^j=|\alpha^j\rangle \langle \beta^j |$ ($|\alpha^j\rangle $ and $|\beta^j\rangle$ are the generic notations for the levels of $j$th atom), $\sigz=\siguu-\sigll$,  $w$ is the incoherent pumping rate, $\gamma$ is the spontaneous rate of the lasing transition, $\gamma_\parallel=w+\gamma$, $\gamma_\perp = (\gamma+w)/2 + \gamma_R$, $\gamma_{R}$ is the incoherent dephasing rate (limited from the bottom by the value $\xi w/2$, where $\xi=\Gamma_1/\Gamma_2$ is the ratio of decay rates $\Gamma_1$ and $\Gamma_2$ of the intermediate pumping state into the lower and the upper lasing states $|l\rangle$ and $|u\rangle$ respectively). $\g$ is the coupling coefficient of the cavity field with the lasing transition in $j$th ion. As is shown in Appendix~\ref{app:I}, 
\begin{equation}
\g = g(\r)= g_0\, e^{ - \frac{\rperp^2}{\w^2}} \, \cos(\k \cdot \r).
\label{equation:27}   
\end{equation}

One may obtain the steady-state (cw) solution of equations~(\ref{equation:24}) -- (\ref{equation:26}) setting time derivatives to zero. Then, from (\ref{equation:24}) follows
\begin{multline}
\aav_{cw}=- \frac{i}{\kappa} \, \frac{3 N}{4 \pi R_c^3} 
\\
\times \int \limits_{-R_c}^{R_c} \int \limits_{0}^{\sqrt{R_c^2-z^2}} \g(\r) \left\langle \sig_{lu}(\g(\r)) \right\rangle_{cw} 2 \pi\, \rperp\, d\rperp\ dz,
\label{equation:28}
\end{multline}
where $R_{c}=N^{1/3} \ell$ is the radius of the Coulomb crystal. In turn, $\left\langle \sig_{lu}(\g(r)) \right\rangle_{cw}=\langle \sig_{lu}^j \rangle_{cw}$ may be expressed via $\aav_{cw}$ with the help of (\ref{equation:25}) and (\ref{equation:26}) as
\begin{equation}
\langle\sig_{lu}(\g(\r))  \rangle_{cw}=
\frac{1}{2} \frac{i\g(\r) \aav_{cw} (w-\gamma)} 
{\gamma_{\perp} \gamma_{\parallel}+|\aav_{cw}|^2\g^2}.
\label{equation:29}
\end{equation}
Substituting (\ref{equation:27}) into (\ref{equation:29}) and (\ref{equation:29}) into (\ref{equation:28}) and reducing $\aav_{cw}$, we obtain the equation
\begin{multline}
1=\frac{(w-\gamma)}{2 \kappa}\frac{3 N}{4 \pi R_c^3}
     \int \limits_{-R_c}^{R_c} 
     \int \limits_{0}^{\sqrt{R_c^2-z^2}} 2 \pi\, \rperp
\\
\times
          \frac{g_0^2 \cos^2(k z) e^{-\frac{2 \rperp^2}{\w^2}} }
           {\gamma_{\perp} \gamma_{\parallel}+|\aav_{cw}|^2 g_0^2 \cos^2(k z)  e^{-\frac{2 \rperp^2}{\w^2}}} 
    \,   d\rperp dz.
\label{equation:30}
\end{multline}
The integral over $\rperp$ may be taken analytically. Then (\ref{equation:30}) transforms into
\begin{multline}
1=\frac{(w-\gamma)}{2 \kappa}\frac{3 N}{4 R_c^3} \frac{w_0^2}{2 |\aav_{cw}|^2} 
\\
\times
     \int \limits_{-R_c}^{R_c} 
     \log\left[\frac{\gamma_{\perp} \gamma_{\parallel}+|\aav_{cw}|^2 g_0^2 \cos^2(k z)}{\gamma_{\perp} \gamma_{\parallel}+|\aav_{cw}|^2 g_0^2 \cos^2(k z) e^{-2 \frac{R_c^2 - z^2}{\w^2}}}\right] dz.
\label{equation:31}
\end{multline}
Because the cavity waist $\w$ and the radius of the crystal $R_c$ are large in comparison with $1/k$, we can average (\ref{equation:31}) on the scale of $2\pi/k$ with the help of relation
\begin{equation}
\frac{2}{\pi}\int\limits_0^{\pi/2} \log (1+b \cos^2 z) d z= 2 \log \left[\frac{\sqrt{1+b}+1}{2} \right].
\label{equation:32}
\end{equation}
It allows to represent (\ref{equation:31}) in the form
\begin{equation}
1=\frac{3 N \zeta^2 (w-\gamma)}{4 \kappa |\aav_{cw}|^2} 
     \left[
        \log\left(1+\sqrt{1+A}\right) - F(A, \zeta) 
     \right],
\label{equation:33}
\end{equation}
where 
\begin{align}
F(A, \zeta) =&  \int\limits_0^1\log\left(1+\sqrt{1+A 
\exp{\frac{2(x^2-1)}{\zeta^2}}}\, \right)dx, \label{equation:34}
\\
A=&\frac{|\aav_{cw}|^2 g_0^2}{\gamma_\parallel \gamma_\perp},
\label{equation:35} \\
\zeta = & \frac{\w}{R_c}. \label{equation:36}
\end{align}
To find the steady-state intracavity field, one has to solve (\ref{equation:33}) numerically.

\section{Prospect for bad-cavity laser with $\rm ^{176}Lu^+$ ions}
\label{sec:Lu}
\begin{figure*}
\begin{center}
\resizebox{0.99\textwidth}{!}
{\includegraphics{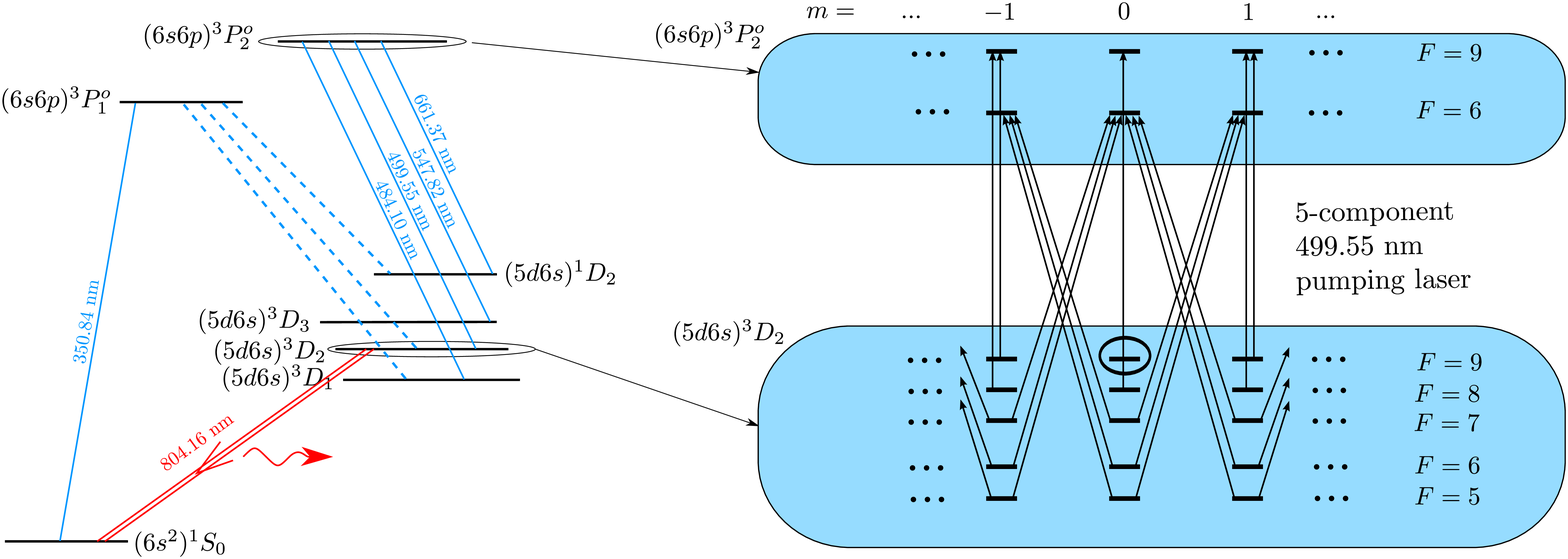}}
\end{center}
\caption{Left: general pumping scheme (hyperfine structure not shown) for a 804 nm bad cavity laser on {$^{176}$Lu} ions. Dashed lines denote the most relevant spontaneous decays, solid lines correspond to both spontaneous and laser-induced transitions (wavelengths are indicated). Right: details of the hyperfine and Zeeman structure of $^3D_2$ and $^3P_2^o$ levels, and transitions induced by 5-component 499.55~nm pumping laser to pump the ions into $|F_u=9, m_u=0\rangle$ upper lasing state.}
\label{fig:Lu}
\end{figure*}
In this section we consider the implementation of a bad-cavity laser using the ${^3D_2} \rightarrow {^1S_0}$ transition in ${\rm ^{176}Lu^+}$ ($I=7$) ion. Briefly this possibility has been mentioned in \cite{Kazakov17}, here we present more detailed quantitative analysis.

A possible pumping scheme is shown in Figure~\ref{fig:Lu}: a 350.84~nm pumping laser populates the $^3P_1^o$ state which decays with a 42~\% probability into the $^3D_2$ upper lasing state, and with a 37.6~\% probability back into the lower lasing state \cite{Paez16}.
The decay of the ${^3P_1^o}$ state will populate also the long-living ${^3D_1}$ state, which can be depopulated via the ${^3P_2^o}$ state with the help of a 484.10~nm laser. Two additional 661.37~nm and 547.82~nm lasers should be applied to pump the atoms out of the  ${^1D_2}$ and  ${^3D_3}$ states populated by the decay of the ${^3P_2^o}$ state. 484.10~nm, 547.82~nm and 661.37~nm lasers may be detuned to the red side and be used also for cooling of the ion ensemble; sympathetic cooling with an additional ion species is also possible.

An important point is that all involved states except the ground state have a hyperfine structure, therefore the pumping lasers should have several frequency components to effectively repump the atoms. Finally, a 5-component 499.55\,nm laser should be employed to pump the populations into the upper lasing state, for example, with specific $F=F_u$ and $m_F=0$. This can be realized if one component of this laser is tuned in resonance with the $|^3D_2, F_e\rangle \rightarrow |^3P_2^o, F_e \rangle$ transition and polarized along the $z$ axis of the trap, coinciding with the direction of the auxiliary magnetic field.

For our calculations, we use values from~\cite{Paez16}, where the spontaneous rate of the ${^3D_2}\rightarrow{^1S_0}$ lasing transition is $\gamma= 4.19 \times 10^{-2}~{\rm s^{-1}}$, the differential scalar polarizability is $\Delta \alpha_0 = -0.9~a_0^3$, and the tensor polarizability of the upper state is $\alpha_{tens}({^3D_2})=-5.6~a_0^3$, where $a_0$ is the Bohr radius.

Let the lower indices $l,u$ and $e$ correspond to the lower lasing ${^1S_0}$, upper lasing ${^3D_2}$, and auxillary $^3P_2^o$ levels. With the help of the $499.55$~nm laser, the populations may be pumped either into one of the $|^3D_2,F_u,m_F=\pm F_u \rangle$ states (both $m_F=\pm F_u$ states may also be populated simultaneously, if $F_u>5$), or into some of the $|^3D_2,F_u,m_F=0 \rangle$ states. 

Quadrupole transitions may, generally speaking, be accompanied by $\Delta m = 0, \pm 1, \pm 2$. In this paper we restrict our consideration to the geometry shown in Fig.~\ref{fig:f1}. In such a configuration, micromotion of the ions takes place primarily in the plane orthogonal to the cavity axis (up to the terms of order of $\epsilon^3$, see (\ref{equation:14}) -- (\ref{equation:15})), and the ions will be confined on length scales significantly smaller than the mode wavelength in axial direction, i.e., in the Lamb-Dicke regime. As shown in Appendix~\ref{app:I}, only the transitions with $\Delta m = \pm 1$ will be coupled with the cavity modes in such a configuration.

Note that two modes ($\sigma^+$- and  $\sigma^{-}$-polarized) with the same eigenfrequency may be excited simultaneously in the cavity. These modes couple the upper $|{^3D_2}, F_u, m_F=0\rangle$ lasing state with two $|{^1S_0}, F_l, m_l=\pm 1 \rangle$ lower states, forming a ``$\Lambda$-system''. Generally speaking, lasing in such a system can not be represented as a simple superposition of 2 independent lasers, because the modes will be coupled via the coherence between two lower lasing levels. However, a detailed investigation of such a system lies beyond the scope of the present paper, and we will consider only a single circularly-polarized mode coupling $|u\rangle = |{^3D_2}, F_u, m_F=0\rangle$ and $|l\rangle = |{^1S_0}, F_l=I, m_F=1\rangle$ states. Note also, that a selective excitation of a single mode may be performed, if the mirrors of the optical cavity will have slightly different transparencies for left- and right-polarized modes, and the lasing threshold will be more easily attainable for one of them. 

\begin{figure*}
\begin{center}
\resizebox{0.95\textwidth}{!}
{\includegraphics{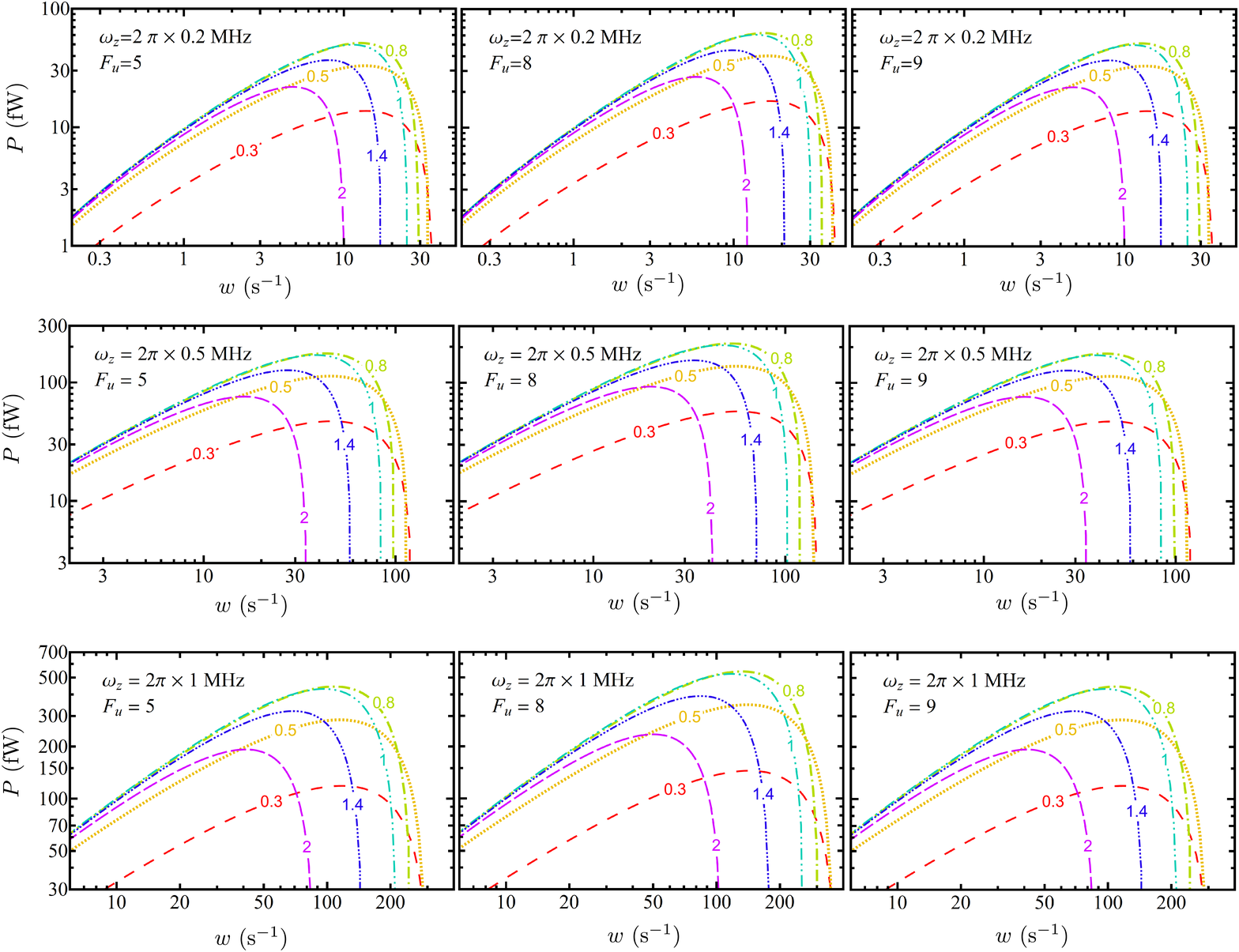}}
\end{center}
\caption{ Output power $P$ for various confinement frequencies $\omega_z$ and upper lasing states $|F_u,m_F=0\rangle$ (on different panels, see labels in upper left corners), and different values of $\zeta=\w/R_c$ (different curves on the same plot, labeled by the values of $\zeta$) as functions of the pumping rate $w$ ($x$ axis) for spherical Coulomb crystal containing $N=10^5$ ${\rm ^{176}Lu^+}$ ions coupled to the cavity with finesse $\F=10^5$. Confinement frequencies are $\omega_z=2\pi \times 0.2$, 0.5 and 1~MHz, which corresponds to the radii $R_c=N^{1/3} \ell$ of the crystal equal to 368, 200 and 126 $\mu$m respectively.}
\label{fig:pow}
\end{figure*}

Using the method presented in Sec.~\ref{sec:micro}, we find that the magic frequency $\Omega_0$ exists for $F_u=5, 8$ and $9$ ($\Omega_0=2\pi \times 25.3$, $45.3$ and $22.3$~MHz respectively). Also, one may find the values of the ``magic magnetic field'' $B_m$, at which the sensitivity of the lasing transition frequency to the fluctuation of this field vanishes in the first order: $B_m= 0.388,$ $-1.035$ and $-1.040$~G for $F_u=5,8$ and 9 respectively.

The maximal value $g_0$ of the coupling coefficient of the Gaussian standing-wave cavity mode with the lasing transition may be estimated as (see Appendix~{\ref{app:I}} for details)
\begin{equation}
g_0 =\Theta_{ul} \sqrt{\frac{5 \pi c^3 \gamma}{\omega^2 \Veff}}.  \label{equation:37}
\end{equation}
where $\Veff=\pi w_0^2 L$ is the effective mode volume, $L$ is the cavity length, $w_0$ is the mode waist,
\begin{equation}
\begin{split}
\Theta_{ul}^2=8(2 F_l+1)&(2J_u+1) \times \\
&\left\{ 
\begin{array}{ccc}
J_l & I & F_l \\
F_u & 2 & J_u
\end{array}
\right\}^2
\left( 
C^{F_u\,m_u}_{F_l\, m_l\, 2\, -1}
\right)^2,  \label{equation:38}
\end{split}
\end{equation}
and $m_u=m_l-1$. It is easy to see that the ratio $g_0^2/\kappa$ does not depend on the cavity length $L$, but only on the cavity finesse $\F$, because $\kappa$ may be expressed via $L$ and $\F$ as $\kappa=\pi c/(\F L)$.

To study the dependence of the output power on the mode waist $\w$, it is convenient to express $\w$ via the radius $R_c$ of the Coulomb crystal as $\w=\zeta R_c$, like in Section~{\ref{Sec:Calc}}. The output power $P$ may be estimated as
\begin{equation}
P=\hbar \omega \kappa\,  |\aav_{cw}|^2, \label{equation:39}
\end{equation}
where $\aav_{cw}$ is the steady-state intracavity field which may be found from the numeric solution of equation (\ref{equation:33}). Note also that $\aav_{cw}$ appears in (\ref{equation:33}) -- (\ref{equation:36}) and (\ref{equation:39}) either as $|\aav_{cw}|^2 g^2$ or as $|\aav_{cw}|^2 \kappa$, therefore, the output power depends on the cavity finesse $\F$, not on the length $L$.

For a quantitative estimation of the output power, we consider a spherical Coulomb crystal containing $N=10^5$ ${\rm ^{176}Lu^+}$ ions, where the lasing transition is one of the $|{^3D_2}, F_u, m_F=0\rangle \rightarrow  \rangle |{^1S_0}, F_l=I, m_F=1\rangle$ quadrupole transitions with $F_u=5$, 8 or 9. The cavity finesse is $\F=10^5$. Also we suppose the repumping efficiency $\xi=0.6$ (i.e., 40 \% of the atoms pumped into the ${^3P^o_1}$ state from the lower lasing state decays back, see Appendix~\ref{app:II} for details). 

Fig.~{\ref{fig:pow}} presents the output power $P$ (\ref{equation:39}) for these 3 transitions for 3 different values of the confinement frequency $\omega_z$ ($\omega_z = 2\pi \times 200~{\rm kHz}$, $500$~kHz and 1~MHz), and different values of $\zeta$. 

\begin{figure}
\begin{center}
\resizebox{0.47\textwidth}{!}
{\includegraphics{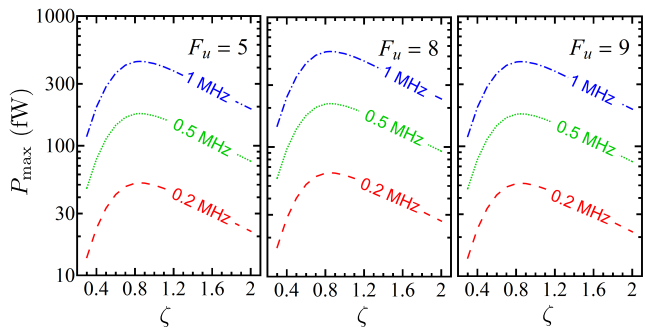}}
\end{center}
\caption{Output power $P_{\rm max}$ optimized with respect to the pumping rate $w$ as a function of $\zeta=\w/R_c$ for different  upper lasing states $|F_u,m_F=0\rangle$ (on different panels, see  labels in upper right corners), and different confinement frequencies $\omega_z$ (labeled curves on the same plot). Parameters are the same as in Fig.~\ref{fig:pow}.}
\label{fig:pmax}
\end{figure}

In figure~\ref{fig:pmax} we show the dependence of the maximum output power $P_{\rm max}$, corresponding to the optimized pumping rate $w$, on the parameter $\zeta$. One can see that the optimal values of $\zeta$ are about 0.8.

Let us discuss the linewidth $\Delta \omega$ of such a bad-cavity laser. Our semiclassical mean-field model can not predict the linewidth; one needs to construct at least some ``second-order theory'' keeping second-order cumulants of the operators related to different ions, as it has been done in \cite{Meiser09, Xu14}, but with a larger amount of groups of ions with different coupling strengths (and shifts, generally speaking). Such a task claims additional attention. However, as an order-of-magnitude estimation, we can take the formula
\begin{equation}
\Delta \omega \approx g^2/\kappa \label{equation:40}
\end{equation}
from \cite{Meiser09}, and substitute $g_0$ for $g$. For $\w=100~{\rm \mu m}$ (which corresponds to $\zeta=0.8$ for spherical Coulomb crystal with $N=10^5$ ions at $\omega_z=2\pi \times 1~{\rm MHz}$) this estimation gives $\Delta \omega \sim 2\pi \times 3 - 4$~mHz for the transitions considered in this section; weaker confinement results in an even narrower linewidth.

Concerning the validity of negligence of the micromotion-induced frequency shift: The optimal values of the repumping rate $w$, at which the maximum output powers are attained, are about 15, 50 and 150~${\rm s^{-1}}$ for $\omega_z=2\pi \times 0.2$~MHz, 0.5~MHz and 1~MHz respectively. These values of $w$ significantly exceed the maximum micromotion-induced shifts $\Delta_{max}$ corresponding to the respective confinement frequencies, see Fig.~\ref{fig:shift}. The repumping rate determines the homogeneous broadening, therefore, near the optimal regime we may neglect the inhomogeneous broadening related to the micromotion-induced shifts, at least for the parameters considered above.

A few words about the quadrupole shift. This shift has been investigated in \cite{Arnold15} for the ${^1S_0} \rightarrow {^3D_1}$ transition in ${\rm Lu^+}$ ion. It has been shown that for $\omega_z=2\pi \times 200$~kHz and a spherical Coulomb crystal with more than a thousand ions, the distribution of the quadrupole shift is symmetric and has a dispersion below 0.1~Hz. The quadrupole shift scales as $\ell^{-3}$, or as $\omega_z^{2}$. Because the quadrupole moment of the ${^3D_1}$ and ${^3D_2}$ states are similar, we estimate that for $\omega_z=2\pi \times 1$~MHz, the dispersion of the quadrupole shifts does not exceed a few Hz, which is much less than the optimal repumping rate. 

Finally, let us compare the Zeeman splitting with the tensor Stark shift. One can easily calculate the Zeeman shifts between the upper lasing state $|F_u,m_F=0\rangle$ and the nearby Zeeman state $|F_u,m_F=1\rangle$ at the respective ``magic'' value of the magnetic field $B_z$: they are $211$, $258$, and $377$ kHz respectively. At the same time, at $\omega_z=2 \pi \times 1$~MHz and $N=10^5$, the tensor Stark shift will be only about 1~kHz on the edge of the Coulomb crystal. Therefore, the tensor Stark shift is small in comparison with the Zeeman shift, and the theory in ref. \cite{Itano00} may be applied.


\section{Discussion and conclusion}
\label{sec:con}
In the present paper, we studied the possibility to create a bad-cavity laser on forbidden transitions in cold ions trapped in a linear Paul trap and forming a large Coulomb crystal. We considered the particular case of a spherical Coulomb crystal of ${\rm ^{176}Lu^+}$ ions, where the $|{^3D_2},F_u, m_F=0\rangle \rightarrow |{^1S_2}, F_l=I, m_F=1\rangle$ transition is coupled to the circularly polarized mode of the high-finesse ($\F=10^5$) optical cavity, whose axis coincides with the trap axis and with the direction of external magnetic field. We showed that $10^5$ ions in the trap with $\omega_z=2\pi \times 1$~MHz may provide about 0.5 picowatts of output power with a 150~$s^{-1}$ repumping rate if the mode waist $\w$ is about 80\% of the crystal radius, i.e., $\w \sim 100~\mu$m.

To increase this power, one could increase the number of ions, increase the frequency $\omega_z$ of the radial confinement, or use an elongated trap. Here we consider the main advantages and disadvantages of these measures in some detail.


First, the ion crystal radius $R_c$ scales as $R_c \propto N^{1/3}$. To keep the value $\zeta=R_c/\w$ near the optimum (about 0.8), we have to increase the cavity waist $\w \propto N^{1/3}$, so the coupling coefficient $g \propto N^{-1/3}$. The total output power $P \propto N^2 g^2$ \cite{Meiser09}, which gives the scaling law $P \propto N^{4/3}$. Also, in larger ion ensembles, the maximum micromotion-related shifts grow with $N$, particularly, $\Delta_{max} \propto N^{2/3}$ at $\Omega=\Omega_0$ (\ref{equation:23}). At the same time, a decrease of $g$ may lead to a decrease of the linewidth $\delta \omega$, as shown in (\ref{equation:40}). We should note, however, that both controlling a large number of ions and fabricating a high-finesse resonator with a large mode waist may be technically challenging.

Second, increasing the confinement frequency $\omega_z$ will lead to a scaling $R_c\propto \omega_z^{-2/3}$, which allows one to increase the coupling coefficient $g \propto \omega_z^{2/3}$, keeping the same $\zeta$, so that the output power scales as $P \propto \omega_z^{4/3}$. At the same time, the maximal micromotion-induced shift $\Delta_{max}$ at $\Omega=\Omega_0$ will scale as $\Delta_{max} \propto \omega_z^{8/3}$, as shown in (\ref{equation:23}). Using the semiclassical model with equal coupling $g$ for all the atoms, but with non-zero frequency shifts (\ref{equation:20}), we found that the maximum output power might be attained at $\omega_z \sim 2 \pi \times 10-20$~MHz (for different transitions) if the other parameters are the same as considered in Section~\ref{sec:Lu}. However, such a large value of $\omega_z$ is much higher than typical ion trap axial frequencies \cite{Arnold15, Landa12}. Moreover, the parameter $\epsilon$ is no longer a small parameter at such large $\omega_z$, and the theory presented in Sections~\ref{sec:micro} and \ref{sec:res} is not valid. Finally, the increasing $g$ may lead to a drastic increase of the linewidth $\Delta \omega$.

Third, the Coulomb crystal in the elongated trap will be less regular than in the spherical one, and the quadrupole shifts may play a more significant, non-negligible role. On the other hand, such a method allows one to pack more ions into the same cavity mode. Such a setup should be designed carefully.

In the present paper, we neglect the excitation of the second circularly polarized mode in the cavity. If such a mode will be excited, it will lead to a reduction of the output power. The picture will become more complex because of interactions of these modes via the coherence between the lower lasing states. A detailed investigation of their interaction will be presented in future work. Here we can note that both the output fields will have different polarizations and frequencies, and the frequency difference will be of the order of a few hundreds of Hz (for ``magic'' magnetic fields). The beat signal between two modes may be used for a stabilization of the magnetic field near its ``magic'' value.

In addition to ${\rm ^{176}Lu^+}$, some other ions with metastable states and negative differential polarizabilities may also be considered as candidates, although it seems to be less straightforward to find a proper repumping scheme. Instead, one may implement a ``passive'' scheme with cavity-enhanced non-linear spectroscopy, similar to the one proposed in \cite{Martin11, Westergaard15}. Such a scheme may also be used for locking the frequency of some slave laser to the optical transition, and this approach does not require pumping of the atoms into the upper lasing state. 


We considered the ``collinear'' configuration, where trap axis, cavity axis and external magnetic field are co-aligned. In such a configuration, the cavity mode will be coupled only with $\Delta m = \pm 1$ quadrupole transitions. To allow the coupling with $\Delta m = 0$ and/or $\Delta m = \pm 2$ transitions, some non-zero angle between the cavity axis and the magnetic field should be introduced. It may be attained, for example, by tilting the cavity axis with respect to magnetic field, co-aligned with the trap axis, or by tilting the magnetic field with respect to the cavity axis coinciding with the trap axis.

In the first case (tilted cavity) the broadening related to the tensor Stark shift (the third term in eq.~(\ref{equation:10})) will be suppressed, as well as in the collinear configuration considered in the main text. However, such a scheme does not allow the use of an elongated trap geometry, and will cause additional problems connected with confinement of the atoms to the Lamb-Dicke regime along the cavity axis. Particularly, our estimations shows that for the parameters of the ions considered in the paper, the amplitude $R_{2,i} \ell$ of the micromotion on the edges of the crystal exceeds the wavelength of the mode. Note that this micromotion-related issue may be of less importance for some long-wavelength transitions, such as ${^2D_{3/2}} \rightarrow {^2S_{1/2}}$ and ${^2D_{5/2}} \rightarrow {^2S_{1/2}}$ transitions in ${\rm Ba^+}$ ions. 

Tilting the magnetic field instead of the cavity axis allows the use elongated traps and to keep the ions in the Lamb-Dicke regime, but requires special measures to suppress the tensor Stark shift. As it has been shown in \cite{Arnold16}, compensation of this shift may be performed with a special choice of the magnetic field.

In summary, we have shown that a bad-cavity laser may be realized on a Coulomb crystal composed of ions with negative differential polarizability of the clock transition. As an example, we considered the $^3D_{2} \rightarrow ^1S_0$ lasing transition in ${\rm ^{176}Lu^+}$, with the ions forming a spherical Coulomb crystal in a linear Paul trap. Such a crystal may provide a route to truly steady-state lasing in the bad cavity regime if the proper continuous cooling and pumping is performed.


\section{Acknowledgments}

We are grateful to Murray Douglas Barrett and John Bollinger for fruitful discussion and valuable comments, and Athreya Shankar for useful remarks. The study has been supported by the EU--FET-Open project 664732 NuClock.

\appendix
\begin{figure}[b]
\begin{center}
\resizebox{0.45\textwidth}{!}
{\includegraphics{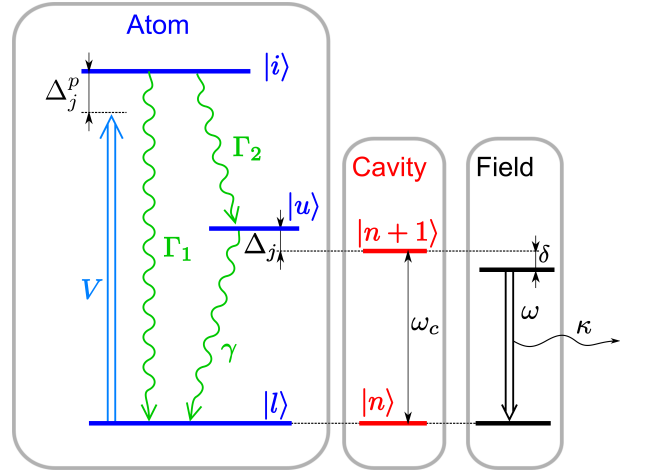}}
\end{center}
\caption{(color online) Level structure and notations used for levels, frequencies, detunings and relaxation rates: $\omega_c$ and $\omega$ are the frequencies of the cavity mode and the output laser field, $\Delta_j$ and $\Delta_j^p$ are the detunings of respective atomic transitions from the output and pumping fields respectively, $\Gamma_1$, $\Gamma_2$ and $\gamma$ are the decay rates.}
\label{fig:3level}
\end{figure}
\section{Semiclassical equations for three-level model and adiabatic elimination of the intermediate state.}
\label{app:II}

Consider the system of $N$ 3-level trapped atoms (ions) with states $|l\rangle$, $|u\rangle$ and $|i\rangle$ , whose $|u\rangle \rightarrow |l\rangle$ transition is coupled with the cavity mode $\a$, and $|l\rangle \rightarrow |i\rangle$ transition is pumped by external field with Rabi frequency $V$, see Fig.~\ref{fig:3level}. Neglecting the dipole-dipole interaction between different atoms and their collective coupling to the bath modes (the role of these effects have been considered in Refs.~\cite{Maier14, Kraemer15}), we can write the master equation for such a system in the form

\begin{equation}
\frac{d\rho}{dt}=-\frac{i}{\hbar}\left[\HH,\hro \right]+ \LL_c\left[\hro\right]
+\sum_j \LL_j\left[\hro\right],
\label{eq:B1}
\end{equation}
where the Lindbladian $\LL_c$ desctribes the relaxation of the cavity field:
\begin{align}
\LL_c[\hro]&=-\frac{\kappa}{2} \left[ \a^+ \a\, \hro+\hro\, \a^+\a-2\, \a\, \hro\, \a^+ \right], \label{eq:B2}
\end{align}
the Lindbladians $\LL_j$ desribes the relaxations of individual atoms
\begin{equation}
\begin{split}
\LL_j[\hro]
&=\frac{\gamma}{2} \left[2 \sig^j_{lu}\hro \sig^j_{ul}-\sig^j_{uu} \hro - \hro \sig^{j}_{uu}\right] \\
&+\frac{\Gamma_1}{2} \left[2 \sig^j_{li}\hro \sig^j_{il}-\sig^j_{ii} \hro - \hro \sig^{j}_{ii}\right] \\
&+\frac{\Gamma_2}{2} \left[2 \sig^j_{ui}\hro \sig^j_{iu}-\sig^j_{ii} \hro - \hro \sig^{j}_{ii}\right], \label{eq:B3}
\end{split}
\end{equation}
and the Hamiltonian in the respective rotating frame has the form

\begin{equation}
\begin{split}
\HH&=\hbar \delta \a^+\a 
+ \hbar \sum_{j} 
   \left( 
       (\Delta_j+\delta) \sig_{uu}^j + \Delta^p_j \sig_{ii}^j
   \right)
\\
&+\frac{\hbar }{2} \sum_{j} \g
   \left(
       \a^+\sig_{lu}^j + \sig_{ul}^j \a
   \right)
 +\frac{\hbar V}{2} \sum_{j} 
   \left(
       \sig_{il}^j + \sig_{li}^j
   \right).
   \label{eq:B4}
\end{split}
\end{equation}
Here and below we use the notation $\sig^j_{rq} = |r^j\rangle\langle q^j|$, where $\g$ is the coupling strength between the $l$th lower and $u$th upper lasing states and the cavity field, $V$ is Rabi frequency of the pumping field.

In the semiclassical (mean-field) approximation, where all the correlators are decoupled, the set of equations for atomic and field expectation values is:
\begin{widetext}
\begin{align}
\frac{d\aav}{dt}  &=-i\left[ \frac{\kappa}{2} + i\delta \right] \aav 
                     -\frac{i}{2}\sum_j \g \siglu, 
\label{eq:B5}
\\
\frac{d\sigll}{dt}&=-i\frac{\g}{2} \left[ \apv \siglu - \aav \sigul  \right]
                   -\frac{iV}{2} \left[ \sigli -\sigil \right]
                   + \gamma \siguu + \Gamma_1 \sigii,
\label{eq:B6}
\\
\frac{d\siguu}{dt}&= i\frac{g}{2} \left[ \apv \siglu - \aav \sigul  \right] 
                   - \gamma \siguu + \Gamma_2 \sigii,
\label{eq:B7}
\\
\frac{d\sigii}{dt}&= \frac{iV}{2} \left[ \sigli -\sigil \right] - \Gamma \sigii
\label{eq:B8}
\\
\frac{d\siglu}{dt}&= -\left(\frac{\gamma}{2} + i (\Delta_j+\delta) \right) \siglu
                     - \frac{i\g}{2} \aav \left[\sigll-\siguu \right]
                     + \frac{iV}{2}\sigiu,
\label{eq:B9}
\\
\frac{d\sigli}{dt}&= -\left( \frac{\Gamma}{2} + i \Delta_j^p \right)\sigli
                     +\frac{i\g}{2} \aav \sigui 
                     - \frac{iV}{2} \left[\sigll - \sigii \right],
\label{eq:B10}
\\
\frac{d\sigui}{dt}&=-\left(\frac{\gamma+\Gamma}{2} +i(\Delta_j^p-\Delta_j-\delta) \right)\sigui
                    +\frac{i\g}{2} \aav \sigli -\frac{iV}{2}\sigul,
\label{eq:B11}                    
\end{align}
\end{widetext}
plus respective equations for $\sigul, \, \sigil,\, \sigiu$. Here $\Gamma=\Gamma_1+\Gamma_2$ is the total decay rate of the intermediate state $|i\rangle$.

Supposing that $\Gamma \gg (V, \Delta_j^p) \gg (\g \aav, \Delta_j, \delta, \gamma)$, we can adiabatically eliminate the intermediate level $|i\rangle$ using  Eqs. (\ref{eq:B9}) -- (\ref{eq:B11}) and corresponding conjugated equations. It gives

\begin{align}
\sigli=\sigil^*&=\frac{-iV}{\Gamma+2i \Delta_j^p} \sigll,
\label{eq:B12}
\\
\sigii&=\frac{V^2}{\Gamma^2+4{\Delta_j^p\,}^2}\sigll,
\label{eq:B13}
\\
\sigui = \sigiu^* &=\frac{-i V}{\Gamma+2i{\Delta_j^p\,}}\sigul.
\label{eq:B14}
\end{align}

Substituting (\ref{eq:B12}) -- (\ref{eq:B14}) into (\ref{eq:B6}) -- (\ref{eq:B8}), we obtain:
\begin{align}
\frac{d \sigz}{dt}&=i\g \left[\apv \siglu - \aav \sigul \right]\nonumber 
\\
                   &+w [1-\sigz] - \gamma[1+\sigz], \label{eq:B15}
\\
\frac{d \siglu}{dt}&=\frac{i\g}{2}\aav \sigz \nonumber 
\\ 
                   &- \left(\frac{\gamma+w}{2}+\gamma_R 
                            +i (\Delta_j + \delta + \Delta_j^{LS}) 
                      \right)\siglu. \label{eq:B16}
\end{align}

Here $\sigz=\siguu-\sigll$,

\begin{equation}
w=\frac{\Gamma_2 V^2}{\Gamma^2+4{\Delta^p_j}^2}
\quad{\rm and} \quad \Delta_j^{LS} = \frac{w (\xi+1)}{\Gamma} \Delta_j^p \label{eq:B17}
\end{equation}
are the incoherent pumping rate and the light shift,

\begin{equation}
 \gamma_R=\xi w/2 \label{eq:B18}
\end{equation}
is the rate of incoherent dephasing caused by the repumping, and $\xi=\Gamma_1/\Gamma_2$.

Let us discuss briefly the light shifts $\Delta_j^{LS}$. First, they are proportional to individual detunings $\Delta_j^p$ of the $|l\rangle \rightarrow |i\rangle$ pumping transition in the $j$th ion from the pumping field. These detunings depends on the micromotion-related second-order Doppler and Stark shifts. If the pumping field is tuned into resonance with the pumping transition of the ion in the center of the trap, they occurs to be proportional to $X_{j,0}^2+Y_{j,0}^2$, as well as the micromotion-related shifts of the lasing transition  (\ref{equation:20}), which may be compensated with the help of the fine tuning of the radio frequency $\Omega$ near its ``magic'' value $\Omega_0$, as mentioned in the end of Section~\ref{sec:res}. Third, light shifts are suppressed, in comparison with the detunings, by a factor of $(\xi+1)w/\Gamma$. For example, in the scheme with ${\rm ^{176}Lu^+}$ ion considered in the paper, $\Gamma=2.8 \times 10^{7}~{\rm s^{-1}}$ and $\xi=0.6$, which for $w \approx 100~{\rm s^{-1}}$ gives $(\xi+1)w/\Gamma\approx 6\times 10^{-6}$. In the present paper, we neglect this shift.
\section{Coupling of electric quadrupole transition with the cavity field}
\label{app:I}
Here we suppose that the cavity mode is a Gaussian standing wave. For the sake of simplicity, we will neglect the beam divergence and the Gouy phase; this simplification is valid, because we only need to calculate the coupling of the cavity mode with ions localized near the cavity waist. Then the electric field of the cavity mode is
\begin{equation}
\EEE(\r)=\e \,\frac{\omega}{c} \sqrt{\frac{8 \pi \hbar c^2}{V_{\rm eff}\, \omega}}\, e^{ - \frac{\rperp^2}{\w^2}}\, \sin(\k \cdot \r) 
      \left[\a + \a^+ \right],
\label{eq:A1}
\end{equation}
where $\e$ is the (complex) polarization unit vector, $\omega$ is the mode frequency, $\k$ is the wave vector, $\w$ is the cavity waist, $\Veff=\pi \w^2 L$ is the effective mode volume, $L$ is the cavity length, $\a$ is the (time-dependent) field operator, and $r_{\perp}=|\r-\k (\k \cdot \r)/k^2|$ is the projection of $\r$ on the plane orthogonal to $\k$. The origin is on the axis of the cavity. If $\w \gg 1/k$, then the interaction of this electric field with quadrupole momentum $\QQ$ of some ion localized in the position $\r$ may be approximately written as
\begin{multline}
\H_{int}=\, \frac{1}{6} \frac{\partial \hat{E}_\alpha}{\partial x_\beta}
         \QQ_{\alpha\beta} 
\approx \,   \frac{{\rm e}_{\alpha} k_{\beta}}{6} 
      \QQ_{\alpha \beta} \left[\a +\a^+ \right] \\
\times \sqrt{\frac{8 \pi \hbar \omega}{V_{\rm eff}}}\, \exp\left[ - \frac{\rperp^2}{\w^2}\right] \, \cos(\k \cdot \r),
\label{eq:A2}
\end{multline}
where the summations over twice appearing Cartesian indices $\alpha, \beta, ...$ are implied here and below, for the sake of brevity. Cartesian components of the quadrupole momentum operator are
\begin{equation}
\QQ_{\alpha \beta} = \int \left(3 x_\alpha x_\beta - \r^2 \delta_{\alpha \beta} \right) \hat{\rho}(\r) d^3 x,
\label{eq:A3}
\end{equation}
where $\hat{\rho}(\r)$ is the operator of the charge density. 

Let us consider some lasing transition between the upper and lower lasing states $|u\rangle$ and $|l\rangle$. Supposing that the ion is placed into the origin, we express the absolute value of the coupling strength $g(\r)$ of the ion situated in $\r$ as
\begin{equation}
g(\r)=
   \frac{2}{\hbar} 
   \left| \left\langle
        l, 1\left| \H_{int} \right|u, 0
    \right\rangle \right| = g_0\, e^{- \frac{\rperp^2}{\w^2}} \, \cos(\k \cdot \r),
\label{eq:A4}   
\end{equation}
where
\begin{equation}
g_0=\sqrt{\frac{8 \pi \omega}{\hbar \, \Veff}} 
    \left|    
        \frac{{\rm e}_{\alpha} k_\beta}{3}
        \left\langle l \left|\QQ_{\alpha \beta} \right|u\right\rangle
    \right|.
\label{eq:A5}                  
\end{equation}
is the coupling coefficient for the ion placed on the cavity axis in the antinode of the mode.

The quadrupole momentum $\QQ$ is the symmetric traceless 2nd rank tensor, and its Cartesian components may be expressed via the spherical components 
\begin{equation}
\QQ_{2q} = \sqrt{\frac{4 \pi}{5}}\int r^2  \hat{\rho}(\r) 
            Y_{2q}\left(\frac{\r}{r}\right) d^3 x
\label{eq:A6}
\end{equation}
as
\begin{align}
\QQ_{xx}=&\sqrt{\frac{3}{2}}\left(\QQ_{22}+\QQ_{2-2} \right)-\QQ_{20}; 
\label{eq:A7}\\
\QQ_{yy}=&-\sqrt{\frac{3}{2}}\left(\QQ_{22}+\QQ_{2-2} \right)-\QQ_{20}; 
\label{eq:A8}\\
\QQ_{zz}=&2 \QQ_{20}; 
\label{eq:A9}\\
\QQ_{xy}=&-i \sqrt{\frac{3}{2}}\left(\QQ_{22}-\QQ_{2-2} \right); 
\label{eq:A10}\\
\QQ_{zx}=& \sqrt{\frac{3}{2}}\left(\QQ_{21}-\QQ_{2-1} \right); 
\label{eq:A11}\\
\QQ_{zy}=&-i \sqrt{\frac{3}{2}}\left(\QQ_{21}+\QQ_{2-1} \right). 
\label{eq:A12}
\end{align}

Now we should express the matrix elements of $\QQ_{2q}$ via the rate $\gamma$ of spontaneous transition. The states $|a\rangle=|\eta_a J_a I F_a m_a\rangle $ ($a=u$ or $l$) are characterized by principal quantum numbers $\eta_a$, the electronic shell angular momenta $J_a$, the nuclear angular momentum $I$, the total angular momenta $F_a$ and its projections $m_a$. Then, according to the well known expression for the electric multipole spontaneous transition rate \cite{Landau4} we can write
\begin{equation}
\gamma = \frac{\omega^5}{15 \hbar c^5} \sum_{F_l,m_l}\left|\left\langle
\eta_u J_u I F_u m_u\big|\QQ_{2q}\big|\eta_l J_l I F_l m_l
\right\rangle\right|^2.
\label{eq:A13}
\end{equation}
Using the Wigner-Eckart theorem \cite{VMK}, we can express the matrix element as
\begin{multline}
\left\langle
\eta_u J_u I F_u m_u\big|\QQ_{2q}\big|\eta_l J_l I F_l m_l
\right\rangle =\\
(-1)^{F_l+J_u+I-2}
\sqrt{2 F_l+1} \, C_{F_{l}m_l 2 q}^{F_u m_u}
\\
\times
\left\{
\begin{array}{ccc}
J_l & I & F_l \\
F_u & 2 & J_u
\end{array}
\right\} 
\left\langle \eta_u J_u ||\QQ_2||\eta_l J_l \right\rangle
\label{eq:A14}
\end{multline}
where $\left\langle \eta_u J_u ||\QQ_2||\eta_l J_l \right\rangle$ is a reduced matrix element. Using the properties of the Clebsch-Gordan coefficients and $6J$-symbols \cite{VMK}
\begin{align}
\sum_{m_l,q}(C_{F_l m_l 2q}^{F_um_u})^2 &=1, 
\label{eq:A15} 
\\
\sum_{F_l} (2 F_l +1) \left\{
\begin{array}{ccc}
J_l & I & F_l \\
F_u & 2 & J_u
\end{array}
\right\}^2 & = \frac{1}{2 J_u+1},
\label{eq:A16} 
\end{align}
we obtain 
\begin{equation}
\left\langle \eta_u J_u ||\QQ_2||\eta_l J_l \right\rangle^2=
\frac{15 \hbar c^5 \gamma (2 J_u+1)}{\omega^5},
\label{eq:A17}
\end{equation}
what gives
\begin{multline}
\left\langle
\eta_u J_u I F_u m_u\big|\QQ_{2q}\big|\eta_l J_l I F_l m_l
\right\rangle^2 =\\
(2 F_l+1)(2 J_u+1) \, (C_{F_{l}m_l 2 q}^{F_u m_u})^2
\\
\times
\left\{
\begin{array}{ccc}
J_l & I & F_l \\
F_u & 2 & J_u
\end{array}
\right\}^2 
\frac{15 \hbar c^5 \gamma}{\omega^5}.
\label{eq:A18}
\end{multline}

Let us calculate the coupling strengths $g$ for circularly polarized cavity mode in configuration shown on the Figure~\ref{fig:f1}, i.e., when $\e=(i \e_y \pm \e_x)/\sqrt{2}$ and $\k=\e_z \omega/c$. Then
\begin{equation}
{\rm e}_{\alpha} k_{\beta} \QQ_{\alpha \beta}= \frac{\omega }{c \sqrt{2}}
\left(i \QQ_{yz}\pm \QQ_{xz} \right)=\frac{\omega \sqrt{3}}{ c} \QQ_{2 \pm 1}.
\label{eq:A21} 
\end{equation}
Substituting (\ref{eq:A21}) into (\ref{eq:A5}), we obtain
\begin{equation}
\begin{split}
g_0 = \frac{\omega}{c \sqrt{3}} \sqrt{\frac{8 \pi \omega}{\hbar \, \Veff}} 
    \left\langle l \left|\QQ_{2\pm 1} \right|u\right\rangle.
\end{split}
\label{eq:A22} 
\end{equation}
One can see that transitions with $m_u-m_l=\pm 1$ can be coupled with the cavity mode in such a configuration.

Let us suppose, for the sake of definiteness, that $m_u=m_l$. Then  
\begin{multline}
g_0^2 = \frac{40 \pi c^3 \gamma }{\Veff\, \omega^2} 
     (2 F_l+1) (2 J_u+1)
\\
    \times 
    \left\{    
    \begin{array}{ccc}  
       J_l & I & F_l \\
       F_u & 2 & J_u
    \end{array}
    \right\}^2
    (C_{F_{l}m_l 2 \pm 1}^{F_u m_u})^2
=\frac{5 \pi c^3 \gamma}{\Veff\, \omega^2} \Theta_{ul}^2,
\label{eq:A23} 
\end{multline}
where $\Theta_{ul}^2$ is given by (\ref{equation:38}).
\section{Hyperfine structure of ${\rm ^{176}Lu^+}$ $^{3}D_2$ state and second-order Zeeman shift}
\label{app:III}

The hyperfine structure of low-lying levels of ${\rm ^{175}Lu^+}$ ion ($I=7/2$) has been measured in \cite{Schueler35}. Particularly, it was found that the energies of the hyperfine sublevels of $^3D_2$ state grows as $F$ increases from 3/2 to 11/2. The distances between adjacent levels are 0.139, 0.210, 0.288 and 0.382 ${\rm cm^{-1}}$. Using the standard expression for the hyperfine energy levels
\begin{align}
\frac{E_{\rm hfs}(F)}{\hbar}&=A_{\rm hfs} \frac{K}{2}\nonumber \\
&+B_{\rm hfs} \frac{\frac{3}{2}K (K+1)-2 I (I+1) J (J+1)}{4 I (2 I-1) J (2 J-1)} \label{eq:C1} \\
{\rm where}& \quad K = F (F+1)-J(J+1)-I(I+1), \label{eq:C2}
\end{align}
we can fit the hyperfine constants as: $A_{hfs,175}=2\pi \times 1935$~{MHz}, $B_{hfs,175}=2\pi \times 1388$~{MHz}.

To estimate the hyperfine constants $A_{hfs,176}$ and $B_{hfs,176}$ for $^{3}D_2$ state of ${\rm ^{176}Lu^+}$ ion, we can use the fact that $B_{\rm hfs}$ is proportional to the nuclear quadrupole moment, and $A_{\rm hfs}$ is proportional to the nuclear g-factor $g_I=-\mu/(I\mu_B)$, where $\mu$ is the nuclear magnetic moment, $\mu_B$ is the Bohr magneton. 

According \cite{Haiduke07}, nuclear quadrupole moments of ${\rm ^{175}Lu^+}$ and ${\rm ^{176}Lu^+}$ are 3415 and 4818 mbarn respectively, which gives $B_{hfs,176}=2 \pi \times 1963$~MHz. In turn, the magnetic moments are $\mu_{175} = 2.2327$ $\mu_N$ and $\mu_{176} = 3.162$ $\mu_N$ according \cite{Stone05}, which gives $A_{hfs,176}=2 \pi \times 1370$~{MHz}.

In the presence of external magnetic field $\B$, the hyperfine energy levels experience a Zeeman splitting. Magnitudes of the Zeeman shift may be found from the diagonalization of the hyperfine-Zeeman Hamiltonian
\begin{multline}
\frac{\hat{H}}{\hbar }=\mu_{B}\left(g_I \II+ g_J \JJ \right)\cdot \B + A_{\rm hfs} \II \cdot \JJ \\
+ \frac{3 B_{\rm hfs}}{4 J (2 J -1) I (2 I -1)} \\
\times \left[ 
   2 (\II \cdot \JJ)^2+\II \cdot \JJ-\frac{2}{3} \II^2 \JJ^2
\right], \label{C3}
\end{multline}
where
\begin{multline}
g_J=g_L \frac{J(J+1)-S(S+1)+L(L+1)}{2 J (J+1)} \\
+g_S \frac{J(J+1)+S(S+1)-L(L+1)}{2 J (J+1)} \label{C4}
\end{multline}
is the electronic Lande g-factor. Taking $g_L=1$, $g_S=2.002319043617$, $L=2$, $S=1$ and $J=2$, we find $g_J\approx 1.16705$. Also, $g_I\approx-0.000246$ for ${\rm ^{176}Lu^+}$.

In weak magnetic field $\B$ the Zeeman shifts $\Delta_{Z|F,m_F=0\rangle}(B)$ of the states $|F,m_F=0\rangle$ are quadratic in  $|\B|$. We find
\begin{align}
\frac{\Delta_{Z|5,0\rangle}(B)}{2 \pi |\B|^2}& =-  440.0~{\rm \frac{Hz}{G^2}},&
\frac{\Delta_{Z|6,0\rangle}(B)}{2 \pi |\B|^2}& =-  15.19~{\rm \frac{Hz}{G^2}} \nonumber \\
\frac{\Delta_{Z|7,0\rangle}(B)}{2 \pi |\B|^2}& =  127.3~{\rm \frac{Hz}{G^2}},&
\frac{\Delta_{Z|8,0\rangle}(B)}{2 \pi |\B|^2}& = 166.3~{\rm \frac{Hz}{G^2}}
\nonumber \\
\frac{\Delta_{Z|9,0\rangle}(B)}{2 \pi |\B|^2}& =  165.5~{\rm \frac{Hz}{G^2}}. \label{C5}
\end{align}

In turn, Zeeman shift $\Delta_{Z,0}$ of the state $|{^1S_0}, m=1\rangle $ (playing the role of the lower lasing state in the scheme considered in the main text) is determined by the nuclear gyromagnetic ratio and is linear in the magnetic field: $\Delta_{Z,0}/B_z=-2 \pi \times 344.3~{\rm Hz/G}$. 

The ``magic'' value $B_m$ of the magnetic field is such a value of the $z$-projection of this field that the frequency difference between the upper- and the lower clock states (or lasing states in our case) is insensitive in the leading order to the fluctuations of this field. Magic fields can be easily found for various transitions from the equating of the first derivatives of the Zeeman shifts of the upper- and the lower lasing state. 


\end{document}